\documentstyle[11pt]{article}

\def\bib{\par\noindent\hangindent=3mm\hangafter=1}
\def\ApJ{Astrophys. J.}
\def\AetA{Astron. Astrophys.}

\begin{document}

\title{\bf New Constraints on the Composition of Jupiter from Galileo
Measurements and Interior Models}
\author{{\sc Tristan Guillot}\thanks{
Work partly done at the Lunar and Planetary Laboratory. Permanent
address: Observatoire de la C\^ote d'Azur, BP 4229, 06304 Nice Cedex
04, France.}\\
Department of Meteorology, University of Reading,\\
Earley Gate, PO Box 243, Reading RG6 6BB, UK\\
E-mail: guillot@met.rdg.ac.uk\\
Tel: +44-118-9875123 ext. 4235\\
Fax: +44-118-9318905\medskip\\
{\sc Daniel Gautier}\\
D\'epartement d'Etudes Spatiales, Observatoire de Paris,\\
5 pl. J. Janssen, 92195 Meudon Cedex, France\medskip\\
{\sc and}\medskip\\
{\sc William B. Hubbard}\\
Lunar and Planetary Laboratory, University of Arizona\\
Tucson, AZ 85721, USA
}
\date{Received 14/04/97; Revised 14/07/97}

\maketitle

\begin{center}
\underline{Submitted to}: {\sl ICARUS}
\end{center}
\bigskip
\bigskip

\noindent Number of manuscript pages: 15 \\
Number of tables: 0 \\
Number of figures: 2 \\

\hphantom{KEYWORDS:} Interiors, planets\\
\hphantom{KEYWORDS:} Chemical abundances

\bigskip\bigskip\noindent 
RUNNING HEAD: Composition of the Interior of Jupiter

\newpage
\setlength{\baselineskip}{22pt}

\begin{abstract}
\bf
\setlength{\baselineskip}{22pt}
Using the helium abundance measured by Galileo in the atmosphere of Jupiter 
and interior models reproducing the
observed external gravitational field, we derive new constraints
on the composition and structure of the planet. We conclude that,
except for helium
which must be  more abundant in the metallic interior than in the molecular 
envelope, Jupiter could be
homogeneous (no core) or could have a central dense core up to
12\,$M_\oplus$. The mass fraction of heavy elements
is less than 7.5 times the solar value in the metallic
envelope and between 1 and 7.2 times solar in the molecular envelope.
The total amount of elements other than hydrogen and helium in the
planet is between 11 and 45\,$M_\oplus$.
\end{abstract}

As shown by Wildt (1938), Jupiter is mainly composed
with hydrogen and helium. However, the question 
of the abundance and partitioning of the 
other species (``heavy elements'') throughout the planet is still unsolved. 
Answering this question would yield
much better constraints on models of formation of the giant planets.
It would also give insight into the behavior of chemical species at
high pressures and generally into physical processes in the
planet's interior.

A precise determination of the seismic properties of the planet would
certainly be the best way to determine accurately its internal
composition. Global oscillations of Jupiter seem to have been 
detected (Mosser et al. 1993), but the accuracy of the measurements
have yet to be improved  and the vibration modes more clearly
identified. To date, constraints on the interior structure of the
planet can be provided solely by interior models matching the
observed gravitational field (i.e. the radius, mass and gravitational
moments $J_2$, $J_4$ and $J_6$) and surface conditions (temperature,
luminosity, atmospheric helium abundance).

Studies of the interior of giant planets (Hubbard \& Marley 1989;
Zharkov \& Gudkova 1992; Chabrier et al. 1992; Guillot et al. 1994b)
have generally been
focused on calculating one possible interior model rather than
calculating the ensemble of all possible models. It is therefore
difficult, from these models, to give constraints on the amount or
distribution of elements in the interior of the planet. Furthermore, the
new determination of the abundance of helium in the upper troposphere  by
the Galileo probe (Niemann et al. 1996) requires a revision 
of these quantities. 

Previous calculations using the Voyager data indicated that:
$Y/(X+Y)=0.18\pm 0.04$ (Gautier et al. 1982; Conrath et al. 1984),
where X and Y are the mass fractions of hydrogen and helium,
respectively. The mass fraction {\em directly} measured by the probe
is now: 
$Y/(X+Y)=0.238\pm 0.007$ (von Zahn \& Hunten 1996). These values have
to be compared to the protosolar value estimated to 
$Y/(X+Y)=0.275\pm 0.005$ (Bahcall and Pinsonneault, 1995).

Since no significant escape of the original hydrogen and helium could
occur over the lifetime of the planet, the interior of the planet must
then be richer in helium than the upper regions in order to insure that the
global hydrogen/helium ratio is equal to the protosolar one. This
conclusion holds for both the Voyager and Galileo
measurements. However, the higher atmospheric helium mass fraction
measured by Galileo results in a smaller abundance of heavy
elements in this region than previously estimated.

We have calculated new optimized Jupiter models using the method
described by Guillot et al. (1994b). We refer the reader to that
article for details on the calculations and references for the
data used. We have used the Galileo
helium/hydrogen ratio, and have included the following uncertainties:
\begin{enumerate}
\item on the equations of state (EOS), for hydrogen, helium and heavy
elements
\item on the temperature profile (convective/radiative)
\item on the internal rotation (solid/differential)
\item on the distribution of helium in the interior.
\end{enumerate}
These uncertainties are discussed in detail below. In addition, we
have included the observational 1-$\sigma$ uncertainties on the
gravitational moments, atmospheric helium abundance and protosolar
helium abundance.

The equation of state of hydrogen and helium is certainly the
most important source of uncertainty in the calculations. A
large part of the jovian interior encompasses a
pressure/tem\-pe\-ra\-tu\-re
region where interactions between particles are important, and 
molecules, atoms, ions and electrons coexist. In such a region, the
behavior of even the simplest molecule, hydrogen, is poorly
understood. It undergoes a phase {\em transition} from a
molecular to a metallic phase at pressures of the order of one to a
few megabar, but it is not known whether this transition is of first order
(i.e. includes a discontinuity of the specific entropy) or not. The
characterization of the two phases is also not obvious:
experimental and theoretical evidence suggest that, at pressures of two
megabars or less, a fraction of the electrons
become itinerant while molecules still exist, before hydrogen
eventually reaches a monoatomic truly metallic state at higher pressures
(of the order of a few megabars). 
Uncertainties in the equation of state are thus
both quantitative and qualitative.

In this work, we choose to estimate these uncertainties by using two
hydrogen-helium EOSs calculated by Saumon, Chabrier \& Van Horn
(1995). The first one ($i$-EOS) assumes that the transition is
continuous and is calculated by smoothly interpolating between the
low-pressure molecular fluid and the high-pressure ionized
plasma. Their second EOS
(PPT-EOS) is consistently calculated and predicts the presence of a
first order plasma phase transition between molecular and metallic
hydrogen, the so-called PPT. This transition occurs around 1.7\,Mbar
for a typical jovian interior profile. In the molecular region, the
densities calculated with these EOSs along isentropes characteristic
of the jovian interior agree with those experimentally determined by
Weir et al. (1996) to better than 2\% (Saumon \& Xie, in
preparation), even though the adiabatic temperature gradients (and hence
temperatures) differ. However, the
electrical conductivities measured indicate a continuous transition
around 1.4\,Mbar (Nellis et al. 1996), and no PPT, at least for
pressures lower than 1.8\,Mbar. On the other hand, Monte-Carlo
simulations by Magro et al. (1996) indicate that a PPT occurs at
slightly larger pressures. We have therefore considered a third case,
using the PPT-EOS, but displacing the PPT to $P=2.5$\,Mbar. Although
the resulting EOS is thermodynamically inconsistent in the
1.7--2.5\,Mbar region, this approach yields a well-behaved density profile
(see Figure~1) and therefore provides another useful estimation of the
uncertainty of the hydrogen-helium EOS.
Finally, we choose to ignore the complexity of the
molecular/metallic hydrogen transition itself, but this is mostly a
semantic issue, assuming that the corresponding uncertainty on the
density profile is properly represented by the different EOSs.

\begin{center}
{\bf Insert Figure~1 here}
\end{center}

The partitioning of elements in Jupiter's atmosphere, and in particular
of helium, represents the second main uncertainty.
There are two mechanisms affecting equilibrium partitioning
of elements in Jupiter's hydrogen.
Stevenson (1982) showed that in the metallic-hydrogen
phase, pressure-ionized elements with $Z > 1$ have limited
solubility.  However, only helium is abundant enough to be directly
affected by such limited solubility (noble gases such as Ne
might be indirectly affected by He-immiscibility -- Roulston \&
Stevenson, 1995). The standard
explanation of Jupiter's (slightly) reduced atmospheric helium abundance
relies on such a phase separation via immiscibility in the metallic phase at
pressures just above the molecular-metallic transition.  

On the other
hand, if a true PPT is present as a first-order phase transition, then
the Gibbs phase rule requires that the concentrations of {\em all} elements
in hydrogen be discontinuous across this boundary. Such partitioning
between the metallic and molecular phases of hydrogen would be in
{\em addition} to any possible phase separation of helium due to
limited miscibility in the metallic phase, and would in principle
affect all elements rather than helium alone.  However, helium and
most heavy elements are expected to be partitioned preferentially
into the molecular phase. The Galileo probe results
for elemental concentrations do not particularly argue
for the predominance of one partitioning mechanism over the other,
but they do confirm that helium is not enriched in the molecular
region.
We emphasize that a composition change due to the PPT or that caused
by limited helium miscibility are expected to occur in the same
pressure region and therefore cannot be distinguished by our models.
Were it not the case (for example if a helium-rich core was formed at
pressures much larger than a few megabars), a slightly wider range of
ice/rock core masses (see below) would be found. 
 
Although it bears little consequences for the structure of the interior
models themselves, our poor knowledge of the equation of state of
heavy elements in Jupiter directly affects the abundance of heavy
elements that is inferred from these models in order to conserve the
protosolar helium/hydrogen ratio. As described by Chabrier et al.
(1992) and Guillot et al. (1994b), the mass fraction of heavy elements
is obtained after optimization of a model using:
\begin{equation}
Z=(Y_{\rm Z}-Y)\frac{\rho_{\rm H}^{-1}-\rho_{\rm He}^{-1}}{
	\rho_{\rm H}^{-1}-\rho_{\rm Z}^{-1}},
\end{equation}
where $Y_{\rm Z}$ and $Y$ are the equivalent and real mass fraction of
helium, respectively, and $\rho_{\rm H}$, $\rho_{\rm He}$ and
$\rho_{\rm Z}$ are the density of pure hydrogen, pure helium, and
heavy elements respectively, along a pressure-temperature profile.
Note that $Z$ is uniform (independent of $T$ and $P$) only in the
limit $\rho_{\rm Z}=\rho_{\rm He}$, or if the ratios $\rho_{\rm
H}/\rho_{\rm Z}$ and $\rho_{\rm H}/\rho_{\rm He}$ are independent of
$T$ and $P$. These variations of $Z$ having no physical meaning, we used a
mass average over the molecular and metallic regions, $Z_{\rm mol}$
and $Z_{\rm met}$, respectively.
In order to estimate the uncertainty on $\rho_Z$, we calculated a low,
a high and an intermediate $\rho_{\rm Z}$
profile. In the case of the low one, we assumed a small mean
molecular weight at low pressures, and interpolated to high pressures
where we used a zero temperature EOS for pure H$_2$O ice adding thermal
effects using Zharkov (1986). The high $\rho_{\rm Z}$ was calculated
using a larger mean molecular weight at low pressures, including
refractory elements (rocks) above 10\,kbar, assuming a 10 times solar
rocks/ices mass mixing ratio and interpolating with zero temperature
EOSs for Olivine (Mg$_2$SiO$_4$) and H$_2$O at high pressures. An
intermediate profile was calculated in the same way, but assuming
solar ices/rocks ratio. At the molecular/metallic transition, $\rho_Z$
was allowed to jump from the low to the high profile, and
vice-versa. The corresponding profiles are shown in
Figure~1 as a hatched region, which represents the uncertainty on
$\rho_Z$, both due to the unknown composition and to the poorly known
EOSs of these elements. 

The possibility that the interior temperature gradient strongly
departs from adiabaticity yields another important source of
uncertainty in the model. The temperature profile retrieved by the
Galileo probe is close to an adiabat at least down to the 20 bar
level (Seiff et al. 1996) but condensation could yield brief subadiabatic or 
superadiabatic events (Guillot 1995).
The possibility that a radiative
region exists in the kbar region (Guillot et al. 1994b) represents the
largest source of uncertainty by far. This was estimated
by calculating Rosseland opacities as in Guillot et al. (1994a), but
using new available opacity data for H$_2$O and H$_2$S (see Marley et
al. 1996). In other regions (except at the PPT) we assumed the
temperature structure to be adiabatic, as any departure from
adiabaticity in the interior (e.g. due to a compositional gradient)
would be indistinguishable from uncertainties in the EOS itself. 

A comparison of the density profiles calculated from the different
equations of state and temperature profiles is given in Figure 1. 
The presence of a radiative region induces an uncertainty of the order
of 5\% on the density in the 10-kbar region of the planet. 
The uncertainty in the H-He EOS is small at both low pressures and at
very high pressures, but reaches as much as 10\% in the intermediate
pressure range of around one to five or so megabars, a pressure range
that involves around a third of the planetary mass.
The $i$-EOS is, on average, significantly less dense than the PPT-EOS,
and therefore requires a larger amount of heavy elements in order to
reproduce the observed mean density of the planet. The presence of a
radiative region has the opposite effect, as it leads to a cooler
planet. 

As in Guillot et al. (1994b), we assumed that a dense central core was
present and included
possible variations in composition (pure ices/pure rocks). Its mass
was allowed to vary in order to fit the observational constraints.
The presence of such a core is important for
constraining formation theories (e.g. Mizuno 1980). 
Although it is often thought that its mass
is constant after the hydrogen-helium envelope has been captured, it
is quite possible that convection redistributes it
towards external regions, therefore yielding a smaller core, if any
(see Stevenson 1985). We emphasize that the presence of a well defined
core is {\em assumed}, and is not a consequence of the observed
gravitational field of the planet. Very similar results would be
obtained by distributing the core over the inner half of the planet.
There is therefore little difference between heavy elements in the
core and in the metallic region.

Finally, a small uncertainty arises from the unknown rotational state
of Jupiter's interior. As shown by Hubbard (1982), assuming that the
planet is differentially rotating on cylinders (with angular velocities
constrained by observations) yields slightly different gravitational
moments $J_2$, $J_4$ and $J_6$ than those inferred from a solid
rotation model. This uncertainty was also included in the present work.

The observed gravitational moments provide no constraints on density
changes which encompass small masses, in particular on the structure
of regions of pressures lower than kbars. Similarly, no real
constraints exist on the structure of the core or its composition
(Hubbard 1989; Guillot et al. 1994b).
Moreover, we find that the resulting models are very {\em insensitive}
to the location of the helium abundance discontinuity, as long as it is
between about 0.5 and 5 Mbar. On the contrary, the models are fairly 
sensitive to
changes in the location of the PPT because it induces significant
changes of the density profile (see Figure~1).

The resulting constraints on the abundance and distribution of heavy
elements in Jupiter are given in Figure 2. The colored surfaces
represent possible solutions for given hydrogen-helium EOSs and
temperature profile. Within each patch
all other uncertainties were included, as discussed previously.
The most significant are: (i) the EOS for heavy
elements; (ii) the value of $J_4$; (iii) the value of $Y_{\rm N}$; and
(iv) the unknown composition of the core. The observational error on
the atmospheric helium abundance measured by the Galileo probe is
small, but it is interesting to note that calculations using the
Voyager value would have yielded values of $Z_{\rm mol}$
larger by 2 or 3 solar values. The difference between adiabatic and
non-adiabatic models is smaller than in Guillot et al. (1994b) because
improved H$_2$O opacities have increased the Rosseland opacities by a
factor $\sim3$ (see the discussion of the problem of hot bands in
Guillot et al. 1994a).

\begin{center}
{\bf Insert Figure~2 here}
\end{center}

A first important result is that the mass of the core can be equal to
zero. In that case, we find it even possible for Jupiter to be
homogeneous ($Z_{\rm mol}=Z_{\rm met}$), if we except small-scale
phenomena and variations of the abundance of helium.
This requires that the density of the metallic phase
is not too large compared to that of the molecular phase ($i$-EOS),
and an important enrichment in heavy elements (4 to 7 times the 
solar value). In the case of the more realistic PPT-EOS, we find that
the presence of a central core is necessary. For all cases, the
core is relatively small: less than 12\,$M_\oplus$ (and less than
8.5\,$M_\oplus$ in the case of a pure rock core). Again, we stress
that the elements that form the core could be distributed over the
inner half of the planet, and that the
distinction between heavy elements in the core and heavy elements in
the metallic region is not obvious.

Other constraints on the amount of heavy elements present in the
planet suffer mainly from our relatively poor knowledge of the
equation of state of hydrogen (and helium) at pressures of around one
to a few megabars. Figure~2 shows that the total mass of
heavy elements (core + metallic + molecular envelope) lies between 11
and 45\,$M_\oplus$. The upper limit corresponds to an enrichment of a
factor 7.4 compared to the solar value, which is quite large. We
estimate that it is probable that the density in the metallic
region lies closer to that predicted by the PPT-EOS than by the
$i$-EOS, as in fact favored by Saumon, Chabrier and Van Horn
(1995). It is thus more likely that $M_{Z\rm tot}\le 33\,M_\oplus$.
Similarly, we obtain that in the metallic region, $Z_{\rm
met}/Z_\odot\le 7.5$, or, when we disregard the $i$-EOS, $Z_{\rm
met}/Z_\odot\le 4.5$.

The constraint found on the mass fraction of heavy elements in the
molecular envelope, $1\le Z_{\rm mol}/Z_\odot\le 7.2$, can be
compared to the measurements of Galileo's mass spectrometer (Niemann
et al., 1996). Those
confirmed that Jupiter's atmosphere is enriched in CH$_4$ by a factor
2.9 compared to the solar value of Anders and Grevesse (1989), and in
H$_2$S by about the same factor. 
The N/H ratio has not yet been evaluated from Galileo, but 
Voyager infrared data suggest an enhancement of 2.2 to 2.4 (Lellouch et al.,
1989; Carlson et al., 1992). The case of H$_2$O is not clear because
it condenses relatively deep in the atmosphere, but it is probably at
least in solar abundance.
Fig. 2 shows that an enrichment in all heavy elements by a factor 2 or 3 
in the molecular region is certainly consistent with all possible interior 
models. 

Reciprocally, it is possible to show from our models that, even though
O is responsible for almost 50\% of the value of $Z_\odot$, large
values of the O/H ratio can be consistent with the observed gravitational
moments. For example, assuming that all other elements are enriched by
a factor 3 in the molecular region, we infer that the enrichment in
H$_2$O can be as high as 12 times the solar value. However,
this requires a relatively hard (low density) H-He EOS in the molecular
region. Using the PPT-EOS only, we find that it is more likely that
H$_2$O is less abundant than 8 times solar in the molecular envelope. 
In any case, these results do not permit to choose between very different
assumptions concerning the enrichment of Jupiter's outer envelope by
meteorites and comets (e.g. Pollack et al. 1986; Owen \& Bar-Nun
1995). It is difficult to infer a lower limit to the abundance of
water in the molecular envelope of the planet, but it appears that it
would be difficult to reconcile an abundance less than solar with the
values of $Z_{\rm mol}$ found, as this would require significant
enrichments in all other elements. 

Most of the uncertainties in the results shown in Figure~2 are due to
the relatively poor knowledge of the equation of state of hydrogen in
the megabar region. Recent improvements in shock-wave experiments (Nellis
et al. 1996) let us hope that a more accurate representation of the
molecular/metallic transition of hydrogen will be achieved in the near
future. These improvements could easily yield abundances of heavy
elements in Jupiter more accurate by 50\%. Similarly, a more precise
value of $J_4$ would lead to better constraints on the structure of
the planet. The optimal precision that should be attained on $J_4$ is
about 3 times the actual one, as one is limited by the
uncertain internal rotation rate of the planet (see Hubbard 1989 for a
discussion). In order to be a useful constraint, $J_6$ should be
measured with a 10 times higher accuracy. Even so, the problem of
differential rotation is even more acute than for $J_4$, and a
detailed study of the internal rotation of the planet should be done
in order for such a measurement to be useful.
Finally, more accurate high pressure opacities and generally a better
understanding of transport processes in the planet are also required
in order to determine precisely the abundance and partitioning of
elements in Jupiter's interior. 

\section*{Acknowledgements}

We thank D. Saumon for helpful discussions, and making available
unpublished calculations, and D.J. Stevenson for constructive referee
comments. 
This work was supported by the European Community through a Training
and Mobility of Researchers grant to T.G. 

\newpage
\section*{References}
\bib{\sc Anders, E., and N. Grevesse} 1989. Abundances of the elements:
meteoritic and solar. {\it Geochim. Cosmochim. Acta} {\bf 53}, 197-214.
\bib{\sc Bahcall, J. N., and M. H. Pinsonneault} 1995.
Solar models with helium and heavy elements diffusion.
{\it Rev. Modern Physics} {\bf 67}, 781--808.
\bib{\sc Chabrier, G., D. Saumon, W. B. Hubbard, and J. I. Lunine} 1992. 
The molecular-metallic transition of hydrogen and the structure of Jupiter
and Saturn. {\it \ApJ} {\bf 391}, 817--826.
\bib{\sc Carlson, B. E., A. A. Lacis, and W. B. Rossow} 1992.
The abundance and distribution of water vapor in the Jovian troposphere 
as inferred from Voyager Iris observations. {\it \ApJ} {\bf 388}, 648-688.
\bib{\sc Conrath, B. J., D. Gautier, R. Hanel, G. Lindal, and A. Marten}
 1984. The helium abundance of Saturn from Voyager measurements. 
{\it \ApJ} {\bf 282}, 807-815.
\bib{\sc Gautier, D., B. B\'ezard, A. Marten, J.-P. Baluteau, N. Scott,
A. Chedin A., V. Kunde, and R. Hanel} 1982. The C/H ratio in Jupiter from the
Voyager infrared investigation. {\it \ApJ} {\bf 257}, 901--912.
\bib{\sc Guillot, T., D. Gautier, G. Chabrier, and B. Mosser} 1994a.
Are the giant planets fully convective? {\it Icarus}, {\bf 112},
337--353.
\bib{\sc Guillot, T., G. Chabrier, P. Morel, and D. Gautier} 1994b.
Non-adiabatic models of Jupiter and Saturn. {\it Icarus}, {\bf 112},
354--367.
\bib{\sc Guillot, T.} 1995. Condensation of methane, ammonia and water and the
inhibition of convection in giant planets. {\it Science} {\bf 269},
1697--1699.
\bib{\sc Hubbard, W. B.} 1982. Effects of differential rotation on the
gravitational figures of Jupiter and Saturn. {\it Icarus} {\bf 52},
509--515.
\bib{\sc Hubbard, W. B.} 1989. Structure and composition of giant planets
interiors. In {\it Origin and Evolution of Planetary and Satellite 
Atmospheres}, S. K. Atreya, J. B. Pollack, and M. S. Matthews, eds.,
University of Arizona Press, Tucson, pp. 539--563.
\bib{\sc Hubbard, W. B., and M. S. Marley} 1989. Optimized Jupiter, Saturn and
Uranus interior models. {\it Icarus} {\bf 78}, 102--118.
\bib{\sc Lellouch., E., P. Drossart, and T. Encrenaz} 1989. A new analysis 
of the Jovian 5-$\mu$m Voyager IRIS spectra. {\it Icarus} {\bf 77},
457--468.
\bib{\sc Magro, W. R., D. M. Ceperley, C. Pierleoni, and B. Bernu}
1996. Molecular dissociation in hot, dense hydrogen. {\it
Phys. Rev. Let.} {\bf 76}, 1240--1243.
\bib{\sc Mizuno, H.} 1980. Formation of the giant planets. {\it
Progr. Theoret. Phys.} {\bf 64}, 544--557.
\bib{\sc Mosser, B., D. Mekarnia, J.-P. Maillard, J. Gay, D. Gautier,
and P. Delache} 1993. Seismological observations with a Fourier transform
spectrometer: detection of Jovian oscillations. {\it \AetA} {\bf 267},
604--622.
\bib{\sc Nellis, W. J., S. T. Weir, and A. C. Mitchell} 1996. Metallization
and electrical-conductivity of hydrogen in Jupiter. {\it Science} {\bf
273}, 936--938.
\bib{\sc Owen, T., and A. Bar-Nun} 1995. Comets, impacts and
atmospheres. {\it Icarus} {\bf 116}, 215--226.
\bib{\sc Pollack, J. B., M. Podolak, P. Bodenheimer and
B. Christofferson} 1986. Planetesimal dissolution in the envelopes of
the forming giant planets. {\it Icarus} {\bf 67}, 409--443.
\bib{\sc Roulston, M. S., and D. J. Stevenson} 1995. Prediction of
neon depletion in Jupiter's atmosphere {\it EOS} {\bf 76}, 343.
\bib{\sc Saumon D., G. Chabrier and H. M. Van Horn} 1995. An equation
of state for low-mass stars and giant planets. {\it
Astrophys. J. Suppl. Ser.} {\bf 99}, 713--741. 
\bib{\sc Seiff, A., D. B. Kirk, T. C. D. Knight, J. D. Mihalov,
R. C. Blanchard, R. E. Young, G. Schubert, U. von Zahn, G. Lehmacher,
F. S. Milos, and J. Wang} 1996. Structure of Jupiter: Galileo Probe 
measurements. {\it Science} {\bf 272}, 844--845.
\bib{\sc Stevenson, D. J.} 1982. Interiors of the giant planets. {\it
Ann. Rev. Earth Planet. Sci} {\bf 10}, 257--295.
\bib{\sc Stevenson, D. J.} 1985. Cosmochemistry and structure of the
giant planets and their satellites. {\it Icarus} {\bf 62}, 4--15.
\bib{\sc Thompson, S. L.} 1990. ANEOS --Analytic Equations of State for
Shock Physics Codes, Sandia Natl. Lab. Doc. SAND89-2951.
\bib{\sc Weir, S. T., A. C. Mitchell, and W. J. Nellis} 1996. Metallization
of fluid molecular-hydrogen at 140 GPA (1.4 Mbar). {\it
Phys. Rev. Lett.} {\bf 76}, 1860-1863.
\bib{\sc Wildt, R.} 1938. On the state of matter in the interior of
planets. {\it Astrophys. J} {\bf 87}, 508--516.
\bib{\sc von Zahn, U., and D. M. Hunten} 1996. The helium mass fraction in
Jupiter's atmosphere. {\it Science} {\bf 272}, 849--851.
\bib{\sc Zharkov, V. N.} 1986. Interior Structure of the Earth and
Planets (Translated by W.B. Hubbard \& R.A. Masteler). Harwood
academic pub., London. 
\bib{\sc Zharkov, V. N., and T. V. Gudkova} 1992. Modern models of giant
planets. In {\it High Pressure Research: Application to Earth and
Planetary Sciences} (Y. Syono and M.H. Manghgnani, Eds.),
pp. 393--401.

\newpage

{\bf Figure 1}:\\
Density profiles in models of Jupiter with a fixed
composition (Y=0.30), using an adiabatic temperature profile and the
$i$-EOS (plain line), adiabatic and PPT-EOS (plain line with
discontinuity), non-adiabatic $i$-EOS (dashed line), and non-adiabatic
PPT-EOS (dotted line). The thin plain and dashed lines correspond to
adiabatic and non-adiabatic 2.5\,Mbar PPT-EOS, respectively (see
text). The temperatures of the models range between 2050 and
2450\,K at a pressure of 10\,kbar, 5300--6300\,K at 1\,Mbar,
10000--13000\,K at 10\,Mbar and 14000--21000\,K at 30\,Mbar.
Upper curves are $T=0$\,K density
profiles for water ice and olivine (from Thompson 1990). The dashed
region represents the assumed uncertainty on the EOS for heavy
elements ($\rho_Z(P,T)$). Within this region, the continuous line
corresponds to our ``preferred'' profile for $\rho_Z$.

{\em Inset}: differences of the decimal logarithm of the Jupiter density
profiles with the same profile using the $i$-EOS and an adiabatic
structure.
The presence of a radiative region leads to a colder and denser model
in the kbar region and deeper. The presence of a PPT yields a denser
metallic region.

\bigskip
{\bf Figure 2}:\\
Constraints on Jupiter's $M_{\rm core}$ (mass of the
central core), $M_{Z\rm tot}$ (total mass of heavy elements), $Z_{\rm
mol}$ (mass fraction of heavy elements in the molecular envelope), and
$Z_{\rm met}$ (mass fraction of heavy elements in the metallic
envelope) from interior models. Each colored surface represents models
calculated with a given hydrogen-helium EOS: PPT-EOS (red),
PPT-EOS with PPT at 2.5\,Mbar (green), $i$-EOS (blue) (see text).
The central regions with a slightly lighter hue have been calculated
with the ``preferred'' heavy elements EOS (see Figure~1), and can be
used to estimate the uncertainty related to the EOS of heavy elements
($\rho_Z$).
Plain and hatched surfaces correspond to adiabatic and non-adiabatic
temperature profiles, respectively. Arrows indicate the magnitude and
direction of uncertainties on $f_{\rm ice}$ (mass fraction of ices in
the core, from 0.5 to 1.0), $Y_{\rm N}$ (mass fraction of helium in
the protosolar nebulae, from 0.275 to 0.28) and $J_4 + \Omega_{\rm
d}$. The latter corresponds to an increase by 1$\sigma$ of $|J_4|$,
and to the assumption that Jupiter rotates on cylinders and not as a
solid planet. The second effect is responsible for about half of the
uncertainty, but only in one direction. For clarity, arrows are
only shown for the PPT-non adiabatic models.
The mass mixing ratios of C, N, O and all other elements (except
H and He) in abundances such that C/H=3$\times$(C/H)$_\odot$,
N/H=3$\times$(N/H)$_\odot$, O/H=3$\times$(O/H)$_\odot$, {\it ...etc}
(the solar abundances are from Anders \& Grevesse 1989) are
shown in the bottom right corner of the figure. The value of $Z$ for
any other composition can be inferred from this ``ruler''.
The molecular envelope represents about 15\% (20\% in the case of the
2.5\,Mbar PPT EOS) of the total mass of the planet ($M_{\rm
Jupiter}=317.83\,M_\oplus$). We used $Z_\odot$=0.0192.

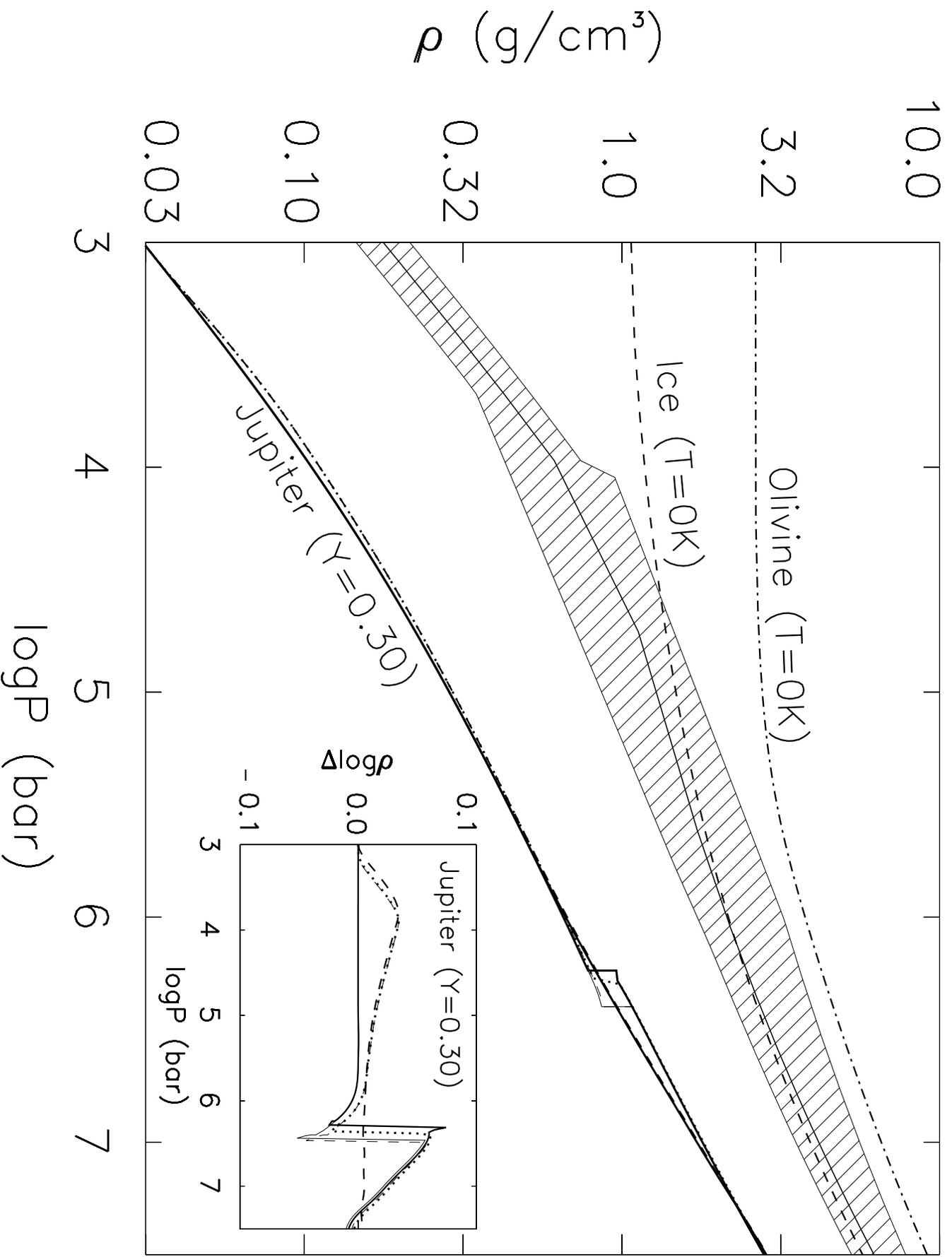
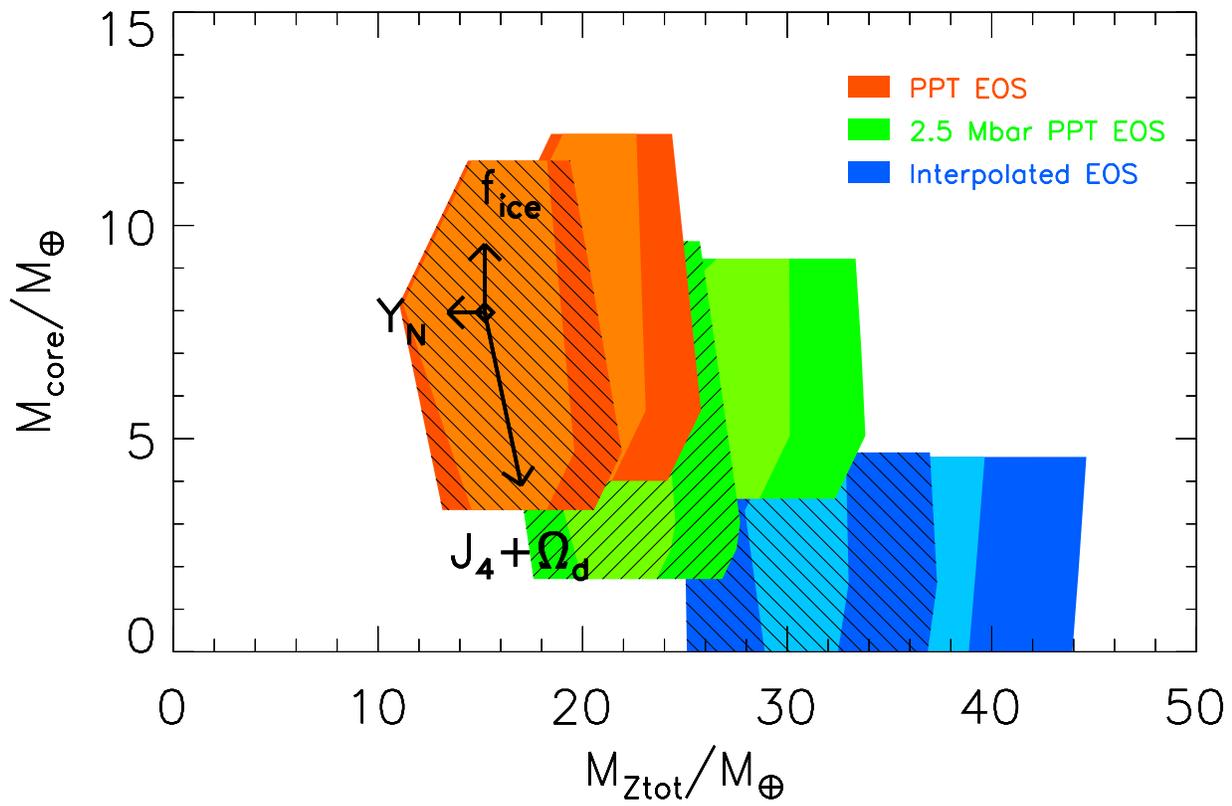
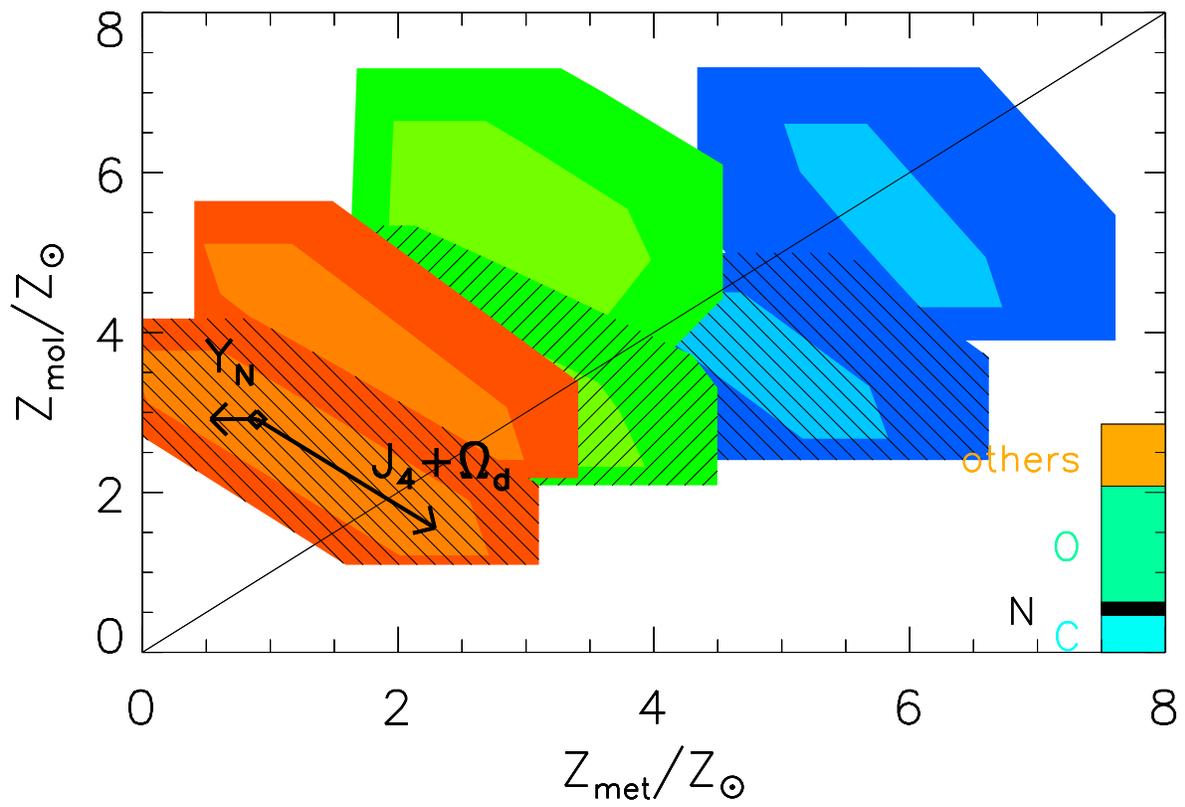
\end{document}